# THE IMPORTANCE OF CONTINUOUS VALUE BASED PROJECT MANAGEMENT IN THE CONTEXT OF REQUIREMENTS ENGINEERING

**Gilbert Fridgen**
FIM Research Center, Augsburg University
gilbert.fridgen@wiwi.uni-augsburg.de

**Julia Heidemann**
McKinsey & Company
julia_heidemann@mckinsey.com

**Motivation**

Companies continuously increased their IT investments over the last decades. Especially the number and complexity of large IT projects is growing. The complexity is intensified by dependencies within one or between different projects and processes and is boosted even further by the growing number of large projects. Another important influence is the rising uncertainty in an increasingly dynamic project management environment. These developments have implications for IT projects and their success. One of the first steps within IT projects is the definition of the project scope including the business case as well as the projects requirements. In that context, modern software development processes and especially methods of agile software developments allow for the ongoing verification and update of these requirements.

However, despite the scientific achievements in the last years also in the context of requirements engineering, there are still a lot of IT projects that fail in the way that they run out of time, budget and value. According to a recent study of the IT Governance Institute about one out of five investments into information technology (IT) is terminated before implementation (ITGI 2011). Flyvbjerg and Budzier (2011) found that one out of six IT projects causes budget deficits of 200% on average. In several cases this can even threaten the existence of the assigning company. The insolvency of the German company Schiesser in 2009 is one of many prominent examples for the economic damages that failing IT-projects can cause: A major reason for the insolvency was the Go-Live of a new enterprise resource planning (ERP)-system leading to erroneous orders for goods (Kessler 2009). Why do companies still fail to achieve the successes initially expected from these IT projects?

Amongst other reasons, unexpected economic risk factors that emerge during the runtime of projects cause budget and time overruns and consequently those high termination rates. Those risk factors lead to the late conclusion that – in contrast to prior expectations – anticipated results cannot be achieved (ITGI 2011). In that context, Flyvbjerg and Budzier found for example that the continuous measurement and controlling of expected projects benefits seems to be positively related to IT project success. What is the case in practice today? If requirements are reconsidered during the runtime of a project, then typically because of technical or cost reasons (e.g. "which features are feasible with the limited budget?"). Financial dependencies between different project parts as well as the measurement of expected projects benefits are mostly neglected today. Moreover, there is a lack of methods to compare the current financial project status with the ex-ante-valuation of the IT project (for example regarding the realized benefits). In many situations, if companies have decided to make a project once, they continue the project even if financial environments have changed.

Requirements engineering and project management methods mostly neglect this challenge so far. Scientific literature primarily focuses on technical aspects or on the financial ex-ante-valuation of IT-projects. However, as stated above empirical studies reveal that a lot of risk factors emerge during the runtime of projects. Therefore, conducting ex-ante-valuations of IT-projects is insufficient. In order to be able to identify emerging risks early and to counteract reasonably, methods for a continuous IT-project-steering are necessary, which as of today to the best of our knowledge are missing within scientific literature.

**The project**

Therefore, we strive to develop an integrated method, which considers costs, benefits, risks, and interdependencies and is, beyond that, easily applicable in practice. For the development of this method, we decided to use an Action Design Research (ADR) approach (Sein et al. 2011). Specific for this research approach is the simultaneous development and the evaluation of an (IT) artifact, which is done in mutual cooperation between practitioners and researchers. Due to the need of companies to evaluate IT projects more holistically and the lack of methods being available and applicable in practice, McKinsey & Company pointed out their need for a methodically sound as well as easy to use method of benefit quantification for a an continuous value based project management for IT projects. Therefore, we developed an approach collaboratively, gathering feedback from practice regarding efficacy and applicability of the method on a regular basis and upholding scientific rigor. Furthermore, we tested the developed method at an industrial client, namely a multinational manufacturing company (in the following referred to as MC), who used the method to evaluate benefits of multiple mobile app development projects. The valuable feedback of both business partners, McKinsey as well as MC, gave us the opportunity to satisfy the criteria of an action design research process and to develop an artifact which fulfills the requirements of all stakeholders from business practice and science and which is currently to be evaluated in real-world project settings.

**Outlook**

The objective should be to further develop methods that allow for a continuous value-based steering of IT-projects (incorporating requirements) across the entire cycle of an IT-project (beginning with ex-ante-valuation over ex-nunc-steering until ex-post-controlling) which incorporates costs, benefits, risks, and interdependencies. By applying such methods for continuous project-steering companies can avoid economic damages caused by failing IT-projects. At the same time, a continuous value-based steering of IT-projects offers the possibility to continuously readjust IT-projects as well as concrete requirements in order to – given underlying circumstances – achieve the highest possible yield.

## Literature

Flyvbjerg, B., Budzier, A.: Why Your IT Project May Be Riskier Than You Think. Harvard Business Review. 89, 23-25 (2011)

ITGI – IT Governance Institute (2011): Global Status Report on the Governance of Enterprise IT (GEIT) – 2011. http://www.isaca.org/Knowledge-Center/Research/Documents/Global-Status-Report-GEIT-10Jan2011-Research.pdf, Access: 15.10.2012

Kessler, G. (2009): Wenn IT voll in die Hose geht. http://wirtschaft.t-online.de/schiesser-wenn-it-voll-in-die-hose-geht/id_47703844/index, Access: 15.10.2012

Sein, M. K., Henfridsson, O., Purao, S., Rossi, M., Lindgren, R.: Action Design Research. MIS Quarterly. 35, 37-56 (2011)